\def\half{\frac{1}{2}}

\def\eq#1{Eq.(\ref{#1})}

\documentclass[amsmath,amssymb,showkeys, showpacs, superscriptaddress]{revtex4}
\usepackage{graphicx}

\begin{document}
\vspace*{0.4 in}
\title{Schr\"odinger spectrum generated by the Cornell potential}
\author{Richard L. Hall}
\email{richard.hall@concordia.ca}
\affiliation{Department of Mathematics and Statistics, Concordia University,
1455 de Maisonneuve Boulevard West, Montr\'eal,
Qu\'ebec, Canada H3G 1M8}

\author{Nasser Saad}
\email{nsaad@upei.ca}
\affiliation{Department of Mathematics and Statistics,
University of Prince Edward Island, 550 University Avenue,
Charlottetown, PEI, Canada C1A 4P3.}

\begin{abstract}
\noindent  The eigenvalues $E_{n\ell}^d(a,c)$ of the $d$-dimensional Schr\"odinger equation with the Cornell potential $V(r)=-a/r+c\,r$, $a,c>0$ are analyzed by means of the envelope method and the asymptotic iteration method (AIM).  Scaling arguments show that it is sufficient to know $E(1,\lambda)$, and the envelope method provides analytic bounds for the equivalent complete set of coupling functions $\lambda(E)$.  Meanwhile the easily-implemented AIM procedure yields highly accurate numerical eigenvalues with little computational effort. 
\end{abstract}

\maketitle
\section{Introduction}\label{intro}
\noindent The Schr\"dinger equation with the Cornell potential  is an important non-relativistic model for the study of quark-antiquark systems \cite{alford,chung, claudio,eichten,eichten1978,eichten1980, evans,chen,hamz}. For example, it is used in describing the masses and decay widths of charmonium states. This Coulomb-plus-linear pair potential was originally proposed for describing quarkonia with heavy quarks \cite{eichten,eichten1978,eichten1980}.  It takes into  account general properties expected from the interquark interaction, namely   Coulombic behavior at short
distances and a linear confining term at long distances \cite{claudio}.  By varying the parameters one can obtain good fits to lattice measurements for the heavy-quark-antiquark static potential \cite{bali}. Although such models have been studied for many years, exact solutions of Schr\"odinger's equation with this potential are unknown. Most of the earlier work either relies on  direct numerical integration of the Schr\"odinger equation or various techniques for approximating the eigenenergies \cite{chung,kang,hall7}.  Without specific reference to a particular physical system, we present a simple and very effective general method for solving Schr\"odinger's equation to any degree of precision in arbitrary dimensional $d>1$.  We write the Cornell potential in the form
\begin{equation}\label{cornell}
V(r)=-\frac{a}{r}+c\,r,
\end{equation}
where $a>0$ is a parameter  representing the Coulomb strength,
and $c>0$ measures the strength of the linear confining term.  The method we use do not require any particular constraint on the potential parameters and thus they are appropriate for any physical problem that may be modelled by this class of potential.  The method of solution is based on a special application of the asymptotic iteration method (AIM, \cite{aim}). AIM is an iterative algorithm originally introduced  to investigate the analytic
and approximate solutions of a second-order linear differential equation of the form
\begin{equation}\label{AIM_Eq}
y''=\lambda_0(r) y'+s_0(r) y,\quad\quad ({}^\prime={d\over dr})
\end{equation}
where $\lambda_0(r)$ and $s_0(r)$ are $C^{\infty}-$differentiable
functions. It states \cite{aim} that: \emph{Given $\lambda_0$ and $s_0$ in
$C^{\infty}(a,b),$ the differential equation (\ref{AIM_Eq}) has the
general solution
\begin{eqnarray}\label{AIM_solution}
\nonumber y(r)= \exp\left(-\int\limits^{r}{s_{n-1}(t)\over \lambda_{n-1}(t)} dt\right) 
 \left[C_2
+C_1\int\limits^{r}\exp\left(\int\limits^{t}\left[\lambda_0(\tau) +
{2s_{n-1}\over \lambda_{n-1}}(\tau)\right] d\tau \right)dt\right]
\end{eqnarray}
if for some $n>0$
\begin{equation}\label{tm_cond}
\delta_n=\lambda_n s_{n-1}-\lambda_{n-1}s_n=0.
\end{equation}
where $\lambda_n$ and $s_n$ are given by
\begin{equation}\label{AIM_seq}
\lambda_{n}=
\lambda_{n-1}^\prime+s_{n-1}+\lambda_0\lambda_{n-1}\hbox{ ~~and~~
} s_{n}=s_{n-1}^\prime+s_0\lambda_{n-1}.
\end{equation}}

\noindent Applications of AIM to a variety of problems have been reported in numerous publications over the past few years. In most applications the functions $\lambda_0(r)$ and $s_0(r)$ are taken to be polynomials or rational functions. However, we show in this paper that the applicability of the method is not restricted to a particular class of differentiable functions. We consider the case where $\lambda_0(r)$ and $s_0(r)$ involve higher transcendental functions, specifically Airy functions. Provided the computer-algebra system employed has sufficient  information about the functions and their derivatives, they present no difficulty. The paper is organized as follows. In section \ref{sec2}, we set up the $d$-dimensional Schr\"odinger equation for the Cornell potential and present some analytical 
spectral bounds based on envelope methods \cite{hall1,hall3,hall4,hall5,hall6}.  In particular we generalize to $d>1$ dimensions an analytical formula, first derived \cite{hall7} for $d=3$, which exhibits 
energy upper and lower bounds for all the discrete eigenvalues of the problem. In section \ref{sec3}, we present an asymptotic solution that allows us to express Schr\"odinger's equation in a form suitable for the application of AIM. In section \ref{sec4}, we apply AIM to the Cornell potential and discuss some of its numerical results, in particular comparisons with the earlier results of Eichten et al. \cite{eichten1978} and the recent work of Chung and Lee \cite{chung}.
\section{Formulation of the problem and analytical estimates in $d$ dimensions}\label{sec2}
\noindent The $d$-dimensional Schr\"odinger equation,
in atomic units $\hbar=2\mu=1$, with a spherically symmetric
potential $V(r)$ can be written as 
\begin{equation}\label{Sch_eq}
\left[-\Delta_d +V(r)\right]\psi(r)=E\psi(r),
\end{equation}
where $\Delta_d$ is the $d$-dimensional Laplacian operator, $d > 1,$  and
$r^2=\sum_{i=1}^d x_i^2$. In order to express (\ref{Sch_eq}) in terms of 
$d$-dimensional spherical coordinates $(r, \theta_1, \theta_2,
\dots, \theta_{d-1})$, we separate variables using
\begin{equation}\label{gs_Sch_eq}
\psi(r)=r^{-(d-1)/2}\,u(r)\, Y_{\ell_1,\dots,\ell_{d-1}}(\theta_1\dots\theta_{d-1}),
\end{equation}
where $Y_{\ell_1,\dots,\ell_{d-1}}(\theta_1\dots\theta_{d-1})$ is a
normalized spherical harmonic \cite{atkin} with characteristic value $\ell(\ell+d-2),$ and $\ell=\ell_1=0, 1, 2, \dots$ (the principal angular-momentum quantum number). One obtains the
radial Schr\"odinger equation as
\begin{eqnarray}\label{gs_Sch_eq1}
&\left[-{d^2\over dr^2}+{(k-1)(k-3)\over
4r^2}+V(r)-E\right] \psi_{n\ell}^{(d)}(r)=0, \\
\nonumber &\int_0^\infty \left\{\psi_{n\ell}^{(d)}(r)\right\}^2dr=1, ~~~
\psi_{n\ell}^{(d)}(0)=0,
\end{eqnarray}
where $k=d+2\ell$.  We assume that the potential $V(r)$ is less singular than the centrifugal term so that
for $(k-1)(k-3)\ne 0$ we have
\begin{equation}\label{ursmall}
u(r)\sim A\,r^{({k-1})/{2}},\quad r\rightarrow 0,\quad {\rm where}~A~{\rm is~a~constant}.
\end{equation}
Since $d>1$ it follows that $k>1,$ and meanwhile $k=3$ only when $\ell=0$ and $d=3.$ 
Thus in the very special case $k=3$, $u(r)\sim Ar$ (as we have for the Hydrogen atom), and we see that \eq{ursmall} is also valid when $k=3.$
We note that the Hamiltonian and the boundary
conditions of \eq{gs_Sch_eq1} are invariant under the transformation
$$(d, \ell)\rightarrow (d\mp2, \ell\pm 1), $$
thus, given any solution for fixed $d$ and $\ell$, we can immediately
generate others for different values of $d$ and $\ell$. Further, the
energy is unchanged if $k=d+2\ell$ and the number of nodes $n$ is
constant: this point has been discussed, for example, by Doren \cite{doren1986}.
 Repeated application of this transformation produces a
large collection of states.   In the present work, we study the $d$-dimension Schr\"odinger eigenproblem
\begin{eqnarray}\label{Sch}
&\left[-{d^2\over dr^2}+{(k-1)(k-3)\over 4r^2}-{a\over r}+c\,r\right]u_{nl}^d(r)=E_{nl}^d u_{nl}^d(r),\\
\nonumber &k=d+2\ell,~a>0, ~ 0<r<\infty, \quad u_{nl}^d(0) = 0.
\end{eqnarray}
 Because of the presence of the linear confining term in the potential, for $c>0$ the spectrum of this problem is entirely discrete: a formal proof for $d>2$ is given in Reed-Simon IV \cite{simon}.
\medskip

If the parametric dependence of the eigenvalues on the potential coefficients $a$ and $c$ is written $E = E(a,c),$
then elementary scaling arguments reduce the dimension of the parameter space to one by means of the equation
\begin{equation}\label{ebounds}
E(a,c) = a^{2}\,E(1,\lambda),\quad {\rm where}\quad \lambda = \frac{c}{a^3}.
\end{equation}
Since $V(r)$ is at once a convex function of $-1/r$ and a concave function of $r^2$, the envelope method 
\cite{hall1,hall3,hall4,hall5,hall6} can be used to derive lower and upper energy bounds based on the comparison theorem and the known exact solutions for the pure Hydrogenic and oscillator problems in $d$ dimensions. It turns out \cite{hall7} that the bounds can be expressed by a formula for $\lambda$ as a function of $E(1,\lambda).$  We have generalized the $d=3$ result of Ref.\kern 2pt \cite{hall7} to $d>1$ dimensions and we obtain:
\begin{equation}\label{eformula}
\lambda = \frac{2\nu^2E^3 -E^2\left[(1+3\nu^2E)^{\half}-1\right]}
{\left[(1+3\nu^2E)^{\half}-1\right]^3}\equiv g(E),\quad E\ge -\frac{1}{4\nu^2},
\end{equation}
which formula yields an upper bound when $\nu = 2n+\ell+d/2$ and a lower bound when $\nu = n+\ell +(d-1)/2.$  It is interesting that
 this entire set of lower and upper (energy) curves are all scaled versions, for example, of the single  ground-state curve.  Again,
 $n = 0,1,2,\dots$ counts the nodes in the radial eigenfunction. Thus by using a computer solve routine to invert the function $g(E)$ in \eq{eformula} for each of the two values of $\nu$, the energy bounds we can be written in the form
\begin{equation}
E(a,c) = a^2g^{-1}_{\nu}(c/a^3).
\end{equation}
For the $s$-states, sharper upper bounds may be obtained (via envelopes of the linear potential) in terms of the zeros of the Airy function.  This is about as far as we can go generally and analytically with this spectral problem.

\section{Asymptotic solution}\label{sec3}
\noindent We note first that the differential equation (\ref{Sch}) has one regular singular
point at $r = 0$ with exponents given by the roots of the indicial equation 
\begin{equation}\label{indicial}
s(s-1)-{1\over 4}(k-1)(k-3)=0,
\end{equation}
and an irregular singular point at $r = \infty$.  For large $r$, the differential equation \eq{Sch} assumes the asymptotic form
\begin{equation}\label{Sch_asy}
\left[-{d^2\over dr^2}+c\,r\right]u_{nl}^d(r)\approx 0
\end{equation}
with a  solution
\begin{equation}\label{asy_sol}
u_{nl}^d(r)\approx Ai\left(c^{1/3}\,r\right),\qquad  u_{nl}^d(\infty)\approx 0,
\end{equation}
where $Ai(z)$ is the well-known Airy function \cite{abr}. Since the roots $s$ of Eq.(\ref{indicial}), namely,
\begin{equation*}
s_1=\half(3-k),\qquad s_2=\half(k-1),
\end{equation*}
determine the behavior of $u_{nl}^d(r)$ as $r$ approaches $0$, only $s>1/2$ is acceptable, since only in this case is the mean value of the kinetic energy finite \cite{landau}.
Thus, the exact solution of (\ref{Sch}) assumes the form
\begin{equation}\label{gen_sol}
u_{nl}^d(r)=r^{(k-1)/2}Ai\left(c^{1/3}\,r\right)~f_n(r),\quad c\neq 0,\quad k=d+2l,
\end{equation}
where we note that $u_{nl}^d(r)\sim r^{(k-1)/2}$ as $r\rightarrow 0$. On  insertion of this ansatz wave function into (\ref{Sch}), we obtain the differential equation for the functions $f_n(r)$ as
\begin{eqnarray}\label{secondorderde}
-r\,f_n''(r)+\left(1-k-2\,r\,\frac{d}{dr}\ln[Ai(c^{1/3}\,r)]\right)f_n'(r)+\left(-a-E\,r-(k-1)\frac{d}{dr}\ln[Ai(c^{1/3}\,r)]
\right)f_n(r)=0.
\end{eqnarray}
\section{Application of the asymptotic iteration method}\label{sec4}
\noindent For arbitrary values of the potential parameters $a$ and $c$, AIM is an effective method to compute the eigenvalues accurately as roots of the termination condition \eq{tm_cond}, which plays a crucial role. The AIM sequences $\lambda_n(r)$ and $s_n(r)$, $n=0,1,\dots$, depend on the (unknown) eigenvalue $E$ and the variable $r$:  thus $\delta_n$ is an implicit function 
of $E$ and $r$. If the eigenvalue problem is analytically solvable, the roots of the termination condition \eq{tm_cond} are independent of the variable $r$ in the sense that the roots of $\delta_n=0$ are independent of any particular value of $r$.  In this case, the eigenvalues are simple zeros of this function. For instance, in the case of a pure Coulomb potential $V(r)=-a/r$, $a>0$, the exact solutions of Sch\"odinger equation
\begin{eqnarray}\label{Sch_Coulomb}
&\left[-{d^2\over dr^2}+{(k-1)(k-3)\over 4r^2}-{a\over r}\right]u_{nl}^d(r)=E_{nl}^d u_{nl}^d(r),\\
\nonumber &k=d+2\ell,~a>0, ~ 0<r<\infty, \quad u_{nl}^d(0) = 0.
\end{eqnarray}
By means of the asymptotic solutions near $r=0$ and $r=\infty$, \eq{Sch_Coulomb} assumes the form
\begin{equation}\label{gen_sol}
u_{nl}^d(r)=r^{(k-1)/2}e^{-\kappa\,r}~f_n(r),\quad k=d+2l,\quad \kappa =\sqrt{-E_n},
\end{equation}
where the functions $f_n$ satisfy the differential equation
\begin{equation}\label{secondorderdeReducd}
\nonumber f_n''(r)=\left(2\kappa+\frac{\left(1-k\right)}{r}\right)f_n'(r)+\frac{(-a+(k-1)\kappa)}{r}f_n(r),
\end{equation}
for $n=0,1,2,\dots$ Thus, continuing the pure Coulomb case, with 
\begin{equation}\label{AIM_Seq_1}
\lambda_0(r)=2\kappa+\frac{\left(1-k\right)}{r},\quad
 s_0(r)=\frac{-a+(k-1)\kappa}{r}
\end{equation}
we  use AIM to compute the sequences $\lambda_n$ and $s_n,~ n=0,1,2,\dots$ initiated with $\lambda_{-1}(r)=1$ and $s_{-1}(r)=0$. The termination condition is $\delta_n=0,n=0,1,2,\dots$ We observe that if $\delta_{n}=0$, then $\delta_{n+1}=0$ for all $n$. Direct computation implies
\begin{eqnarray*}
\delta_0=0,& \quad E_0=-\frac{a^2}{(k-1)^2}\\
\delta_1=0,&\quad E_0=-\frac{a^2}{(k-1)^2},\quad  E_1=-\frac{a^2}{(k+1)^2}\\
\delta_2=0,&\quad E_0=-\frac{a^2}{(k-1)^2},\quad  E_1=-\frac{a^2}{(k+1)^2},\quad  E_2=-\frac{a^2}{(k+3)^2}\\
\delta_3=0,&\quad E_0=-\frac{a^2}{(k-1)^2},\quad  E_1=-\frac{a^2}{(k+1)^2},\quad  E_2=-\frac{a^2}{(k+3)^2},
 E_3=-\frac{a^2}{(k+5)^2}\\
\end{eqnarray*}
and in general
\begin{eqnarray*}
\delta_n&=0\quad\Longrightarrow\quad  E_j=-\frac{a^2}{(k+2j-1)^2},\quad j=0,1,2,\dots, n.
\end{eqnarray*}
as the well-know eigenvalue formula for the Coulomb potential in $d$-dimensions. The situation is quite different in the case of $c\neq 0$.  Here we use AIM with (see equation \eq{secondorderde})
\begin{eqnarray}\label{AIM_Seq_g}
\lambda_0(r)&=&\frac{(1-k)}{r}-2\frac{d}{dr}\ln[Ai(c^{1/3}\,r)],\\
s_0(r)&=&-E-\frac{a}{r}-\frac{(k-1)}{r}\frac{d}{dr}\ln[Ai(c^{1/3}\,r)],
\end{eqnarray}
where the termination condition $\delta_n=0$ is a function of both $r$ and $E$, namely
\begin{equation}\label{deltafunction}
\delta_n\equiv \delta_n(E;r)=0.
\end{equation}
The  problem is then finding an initial value $r=r_0$ that would stabilize the recursive computation of the roots by the  termination condition \eq{deltafunction} for all $n$. This is still an open problem  with no general strategy to locate this initial value. A good choice for $r_0$ depends on the shape of the potential under consideration and sometimes on the asymptotic solution process itself.  Thus two policies for the choice  of $r_0$ are:  (1) the point where the minimum of the potential  occurs if it is not infinity; (2) the point where the maximum of the ground-state asymptotic solution occurs. 
 For the Cornell potential, because of the attractive Coulomb term,  the potential function is not bounded below and we therefore choose $r_0$ to be the location of the maximum of the ground-state wave function as follows. The asymptotic solution is given by:
\begin{equation}\label{asy_sol}
u_{\rm as}(r)\approx r^{(k-1)/2}Ai\left(c^{1/3}\,r\right),
\end{equation}
and we suppose that $\hat{r}$ is the position of the maximum of $u_{\rm as}(r).$  
We start with $r_0= \hat{r},$ then 
we gradually increase the value of $r_0$ until we reach stability in the computational process, in the sense that it converges in few iterations. Thus, once a suitable value is found for $r_0$ for a parameter patch,  the actual eigenvalue calculations are extremely fast. We only found one difficulty with this approach for the present problem, namely  when $c$ is small so that the wave function is very spread out (like the pure Coulomb case).  In order to deal wih this, we adopted the following strategy: we took $r_0$ as a point at which the tail of the asymptotic solution
 \eq{asy_sol} starts to diminish rapidly.  In Figure 1, we show plots of $u_{\rm as}$ for different values of $c$. These graphs suggest that the starting value of $r_0=20$ for the potential $V(r)=-1/r+0.01\,r$, the starting value of $r_0=5$ for the potential $V(r)=-1/r+r$, and $r_0=1$ for $V(r)=-1/r+100\,r$. 
\begin{figure}[!h]
\centering
\includegraphics[width=5cm, height=3cm]{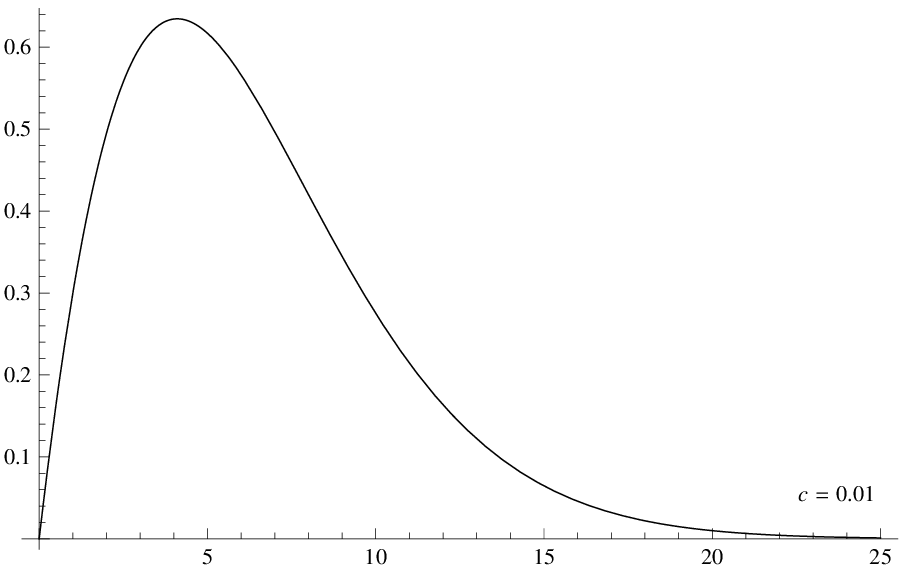}
\includegraphics[width=5cm, height=3cm]{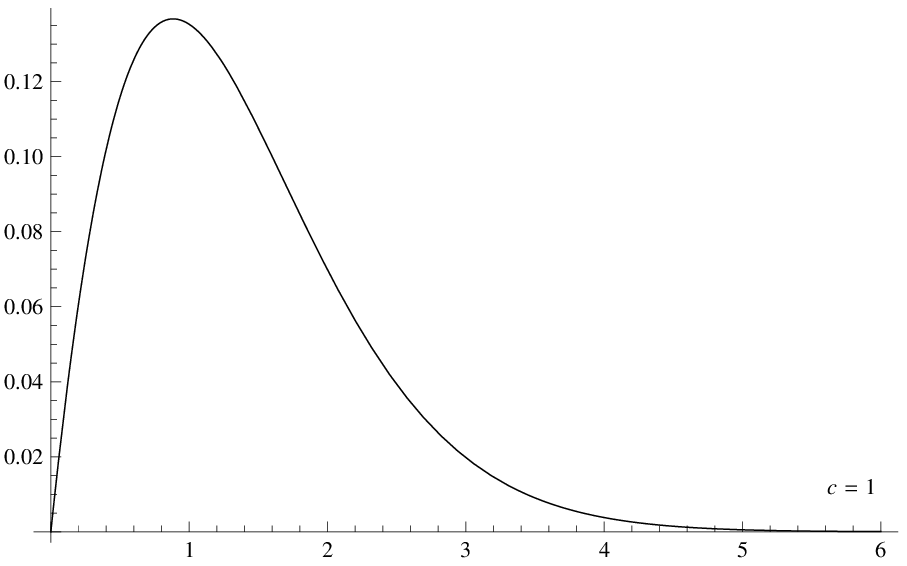}
\includegraphics[width=5cm, height=3cm]{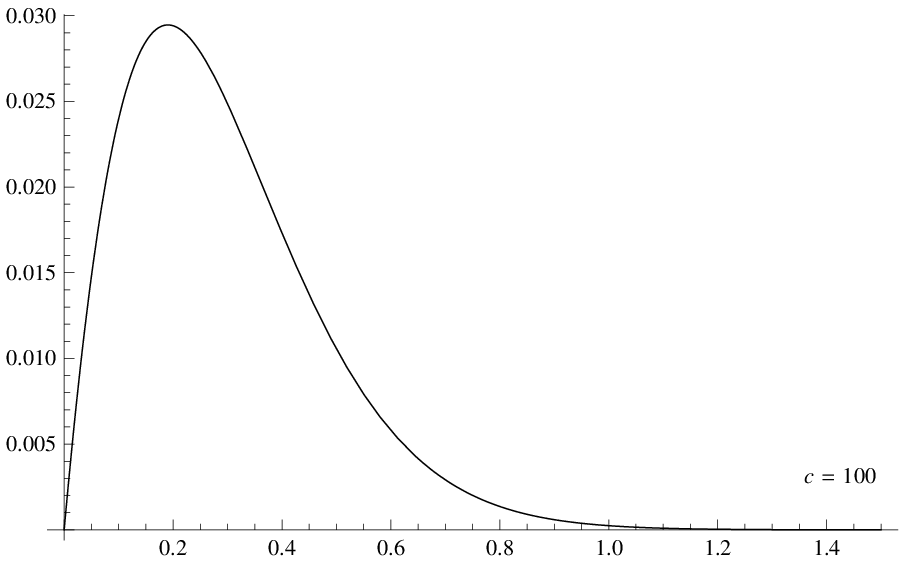}
\caption{The spatial spread of the asymptotic solution $u_{\rm as}$ as $c$ increases.}\label{Fig1}
\end{figure}
For the purpose of consistency we  have calculated each eigenvalue to 12 significant figures and recorded in a subscript the minimum number of iterations required to reach this precision. The computation of the Airy function is 
straightforward, thanks to Maple, where the \emph{`AiryAi'} and its derivative are built-in functions. The eigenvalues reported in Table \ref{table:tab1}  were computed using Maple version 16 running on an Apple iMAC computer in a high-precision environment.  In order to accelerate our computation we have written our own code for a root-finding algorithm instead of using the default procedure {\tt Solve} of \emph{Maple 16}. The results of AIM may be obtained to any desired degree of precision: we have reported most of our results to twelve decimal places, and those of Table 
\ref{table:tab3} to fifteen places, as an illustration.  Of course, once the energy eigenvalue has been determined accurately, it is straightforward to integrate \eq{Sch} to find the corresponding wave function $u(r):$
we exhibit the result in Fig. \ref{Fig2}.

\begin{figure}[ht]
\centering
\includegraphics[scale=0.5]{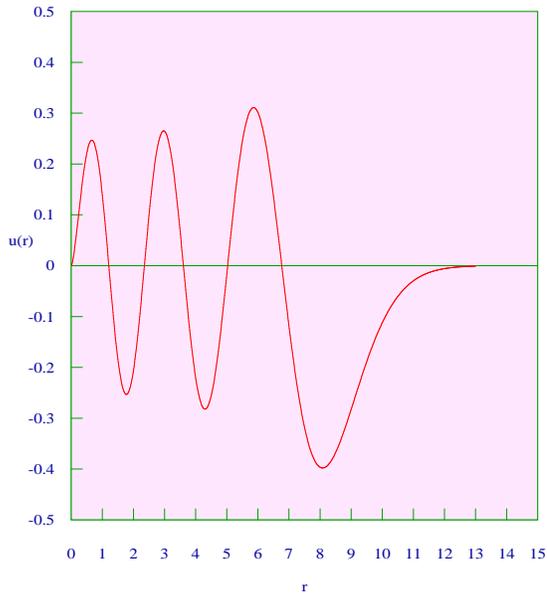} 
\caption{ The wave function $u(r)$ obtained by integrating \eq{Sch} with $ k = 4$  and
the energy eigenvalue $E = E^{d}_{n\ell} = 8.997414071$ taken from Table \ref{table:tab1}.  This corresponds, for example, to
the case $d = 4, \ell = 0,$ and $n = 6.$}
\label{Fig2}
\end{figure}

\begin{table}[h] \caption{Eigenvalues $E_{nl}^{d=3,4}$ for $V(r)=-1/r+r$. The initial value used by AIM is $r_0=5$. The subscript $N$ refers to the number of iteration used by AIM.\\ } 
\centering 
\begin{tabular}{|c|c |p{1.4in}||c|c| p{1.5in}|}
\hline
$\ell$&$n$&$E_{n0}^{d=3}$&$\ell$&$n$&$E_{0l}^{d=3}$\\ \hline
$0$&0&$~1.397~875~641~660_{N=58}$&$0$&0&$~1.397~875~641~660_{N=70}$\\
~&1&$~3.475~086~545~396_{N=73}$&$1$&~&$~2.825~646~640~704_{N=56}$\\
~&2&$~5.032~914~359~536_{N=73}$&$2$&~&$~3.850~580~006~803_{N=51}$\\
~&3&$~6.370~149~125~486_{N=72}$&$3$&~& $~4.726~752~007~096_{N=43}$ \\
~&4&$~7.574~932~640~591_{N=66}$&$4$&~&$~5.516~979~644~329_{N=37}$\\
~&5&$~8.687~914~590~401_{N=82}$&$5$&~&$~6.248~395~598~411_{N=33}$\\
\hline
\hline
$\ell$&$n$&$E_{n0}^{d=4}$&$\ell$&$n$&$E_{0\ell}^{d=4}$\\ \hline
$0$&0&$~2.202~884~354~411_{N=56}$&$0$&0&$~2.202~884~354~411_{N=56}$\\
~&1&$~3.998~899~718~709_{N=67}$&$1$& ~ &$~3.363~722~259~378_{N=54}$\\
~&2&$~5.457~656~703~862_{N=68}$&$2$&~&$~4.301~971~630~406_{N=48}$\\
~&3&$~6.740~670~678~009_{N=67}$&$3$&~& $~5.130~492~519~711_{N=41}$ \\
~&4&$~7.909~993~263~956_{N=63}$&$4$&~&$~7.085~515~480~564_{N=37}$\\
~&5&$~8.997~414~071~258_{N=58}$&$5$&~&$~8.799~435~022~938_{N=41}$\\
\hline
\hline
\end{tabular}
\label{table:tab1}
\end{table}
\vskip0.1true in
\noindent In table \ref{table:tab2} we report the eigenvalues for the Schr\"odinger equation with the potential $V(r)=-1/r+0.01\,r$. The AIM iterations used $r_0=20$. In table \ref{table:tab3}, we report the eigenvalues for the Schr\"odinger equation with the potential $V(r)=-1/r+100\,r$ where with $r_0=1$. In Table \ref{table:tab4} we compare our AIM ground-state eigenenergies for the potential $V(r)=-a/r+r$ and different values of the parameter $a$, with those computed earlier by Eichten et al. \cite{eichten1978} using an interpolation technique and that
of Chung and Lee \cite{chung} using the Crank-Nicholson method. Since the asymptotic solution  \eq{asy_sol} is independent of the Coulombic parameter  $a$ we use AIM with $r_0=1$, as shown  in Figure \ref{Fig1}.

\begin{table}[h] \caption{Eigenvalues $E_{nl}^{d=3,4}$ for $V(r)=-1/r+0.01\,r$. The initial value used by AIM is $r_0=20$ or as indicated. The subscript $N$ refers to the number of iteration used by AIM.\\ } 
\centering 
\begin{tabular}{|c|c |p{2in}||c|c| p{2.0in}|}
\hline
$\ell$&$n$&$E_{n0}^{d=3}$&$\ell$&$n$&$E_{0l}^{d=3}$\\ \hline
$0$&0&$~-0.221~030~563~404_{N=79}$&$0$&0&$~-0.221~030~563~404_{N=79}$\\
~&1&$~~~~~0.034~722~241~998_{N=70}$&$1$&~&$~~~~~0.017~400~552~510_{N=61}$\\
~&2&$~~~~~0.141~913~022~811_{N=66}$&$2$&~&$~~~~~0.102~472~150~415_{N=47}$\\
~&3&$~~~~~0.220~287~171~811_{N=60}$&$3$&~& $~~~~~0.159~830~894~613_{N=39}$ \\
~&4&$~~~~~0.344~602~792~592_{N=75}$&$4$&~&$~~~~~0.206~238~109~687_{N=41}$\\
~&5&$~~~~~0.448~055~673~514_{N=85}$&$5$&~&$~~~~~0.246~682~072~100_{N=34}$\\
\hline
\hline
$\ell$&$n$&$E_{n0}^{d=4}$&$\ell$&$n$&$E_{0\ell}^{d=4}$\\ \hline
$0$&0&$~-0.057~503~250~143_{N=69}$&$0$&0&$~-0.057~503~250~143_{N=69}$\\
~&1&$~~~~~0.087~181~857~064_{N=63}$&$1$& ~ &$~~~~~0.065~687~904~463_{N=54}$\\
~&2&$~~~~~0.176~559~165~345_{N=72}$&$2$&~&$~~~~~0.133~067~612~356_{N=43}$\\
~&3&$~~~~~0.247~865~703~619_{N=67}$&$3$&~& $~~~~~0.183~984~697~123_{N=36}$ \\
~&4&$~~~~~0.309~777~243~695_{N=69,r_0=25}$&$4$&~&$~~~~~0.227~037~524~190_{N=37,r_0=25}$\\
~&5&$~~~~~0.365~723~900~484_{N=71,r_0=25}$&$5$&~&$~~~~~0.287~224~084~341_{N=39,r_0=25}$\\
\hline
\hline
\end{tabular}
\label{table:tab2}
\end{table}

\begin{table}[h] \caption{Eigenvalues $E_{nl}^{d=3}$ for $V(r)=-1/r+100\,r$. The initial value used by AIM is $r_0=1$ or as indicated. The subscript $N$ refers to the number of iteration used by AIM.\\ } 
\centering 
\begin{tabular}{|c|c |p{1.7in}||c|c| p{2.0in}|}
\hline
$\ell$&$n$&$E_{n0}^{d=3}$&$\ell$&$n$&$E_{0l}^{d=3}$\\ \hline
$0$&0&~~$46.402~258~652~779_{N=104}$&$0$&0&$~~46.402~258~652~779_{N=75}$\\
~&1&~~$85.339~271~687~574_{N=106}$&$1$&~&$~~70.016~058~921~076_{N=62}$\\
~&2&~$116.728~692~980~119_{N=103}$&$2$&~&$~~89.715~370~910~984_{N=51}$\\
~&3&$~144.315~456~241~781_{N=99}$&$3$&~& $~107.334~329~106~273_{N=46}$ \\
~&4&$~169.460~543~870~657_{N=102}$&$4$&~&$~123.561~985~764~157_{N=56,r_0=1.5}$\\
~&5&$~192.850~291~861~086_{N=103}$&$5$&~&$~138.761~138~633~388_{N=50,r_0=1.5}$\\
\hline
\hline

\end{tabular}
\label{table:tab3}
\end{table}

\begin{table}[h]
\caption{A comparison between the eigenvalues the S-wave heavy quarkonium results of Eichten et al.\cite{eichten1978}, Chung and Lee \cite{chung} and those of the present work,  $E_{00}^{d=3}$ for ground state with the Coulombic parameter $a$ in the potential $V(r)=-a/r+r$. The initial value used by AIM was fixed at $r_0=6$. The subscript $N$ refers to the number of iteration used by AIM.\\ }
\centering
\begin{tabular}{|c|c|c|}\hline
 $a$& $E_{00}^3$ (Eichten et al.\cite{eichten1978}) & $E_{0,0}^3$(AIM)  \\ \hline
$0.2$ &  $2.167~316$ & $2.167~316~208~772~717_{N=104}$ \\ \hline
$0.4$ &  $1.988~504$ & $1.988~503~899~750~869_{N=105}$ \\ \hline
$0.6$ &  $1.801~074$ & $1.801~073~805~646~947_{N=104}$ \\ \hline
$0.8$ &  $1.604~410$ & $1.604~408~543~236~585_{N=103}$ \\ \hline
$1.0$ &  $1.397~877$ & $1.397~875~641~659~907_{N=102}$ \\ \hline
$1.2$ &  $1.180~836$ & $1.180~833~939~744~787_{N=109}$  \\ \hline
$1.4$ &  $0.952~644$ & $0.952~640~495~218~560_{N=110}$  \\ \hline
$1.6$ &  $0.712~662$ & $0.712~657~680~461~034_{N=115}$  \\ \hline
$1.8$ &  $0.460~266$ & $0.460~260~113~873~608_{N=117}$   \\ \hline
\end{tabular}
\vspace{ 0.3 in}

\begin{tabular}{|c||c||c|}\hline
$a$& $E_{00}^3$ (Chung and Lee \cite{chung}) & $E_{0,0}^3$ (AIM) \\ \hline
$0.1$ &$2.253~678$ & $2.253~678~098~810~761_{104}$\\ \hline
$0.3$ &$2.078~949$ &  $2.078~949~440~194~840_{105}$\\ \hline
$0.5$ &$1.895~904$ & $1.895~904~238~476~994_{106}$\\ \hline
$0.7$ &$1.703~935$& $1.703~934~818~031~980_{104}$\\ \hline
$0.9$ &$1.502~415$& $1.502~415~495~453~739_{99~}$ \\ \hline
$1.1$ & $1.290~709$&$1.290~708~615~983~606_{105}$ \\ \hline
$1.3$ &$1.068~171$ &$1.068~171~244~486~971_{109}$ \\ \hline
$1.5$ & $0.834~162$& $0.834~162~211~049~953_{111}$ \\ \hline
$1.7$ & $0.588~049$&$0.588~049~168~557~953_{115}$  \\ \hline

\end{tabular}
\label{table:tab4}
\end{table}
\vskip0.1true in
\section{Conclusion}

\noindent The solution procedure  presented in this paper is based on the asymptotic iteration method
 and is very simple. It yields highly accurate eigenvalues with little computational effort. To our knowledge, this work is the first attempt to employ  the asymptotic iteration method where the AIM sequences $\lambda_n$ and $s_n,  n=0,1,2,\dots$, are computed in terms of higher transcendental functions, rather than polynomials or rational functions.  This simple and practical method can easily be implemented with any available symbolic mathematical software to elucidate the dependence of the energy spectrum on potential parameters. Once accurate eigenvalues are at hand, it is straightforward to obtain the corresponding wave functions

\vskip0.1true in 
\section{Acknowledgments}
\medskip
\noindent Partial financial support of this work under Grant Nos. GP3438 and GP249507 from the 
Natural Sciences and Engineering Research Council of Canada
 is gratefully acknowledged by us (respectively RLH and NS). 


\medskip

\section*{References}
\begin{thebibliography}{00} 
\bibitem{alford} J. Alford and M. Strickland, \emph{Charmonia and bottomonia in a magnetic field}, Phys. Rev. D \textbf{88} (2013) 105017.

\bibitem{chung} H-S Chung and J. Lee, \emph{Cornell potential parameters for S-Wave heavy quarkonia}, J. Korean Phys. Soc., \textbf{52} (2008) 1151.

\bibitem{eichten}  E. Eichten, K. Gottfried, T. Kinoshita, J. Kogut, K. D. Lane, and T.-M. Yan, \emph{Spectrum of Charmed Quark-Antiquark Bound States}, Phys. Rev. lett. \textbf{34} (1975) 369
[Erratum-ibid. 36, 1276 (1976)].

\bibitem{eichten1978}  E. Eichten, K. Gottfried, T. Kinoshita, J. Kogut, K. D. Lane, and T.-M. Yan, \emph{Charmonium: the model}, Phys. Rev. D \textbf{17} (1978) 3090.

\bibitem{eichten1980}  E. Eichten, K. Gottfried, T. Kinoshita, J. Kogut, K. D. Lane, and T.-M. Yan, \emph{Charmonium: comparison with experiment}, Phys. Rev. D \textbf{21} (1980) 203.


\bibitem{evans} P. W. M. Evans, C. R. Allton, and J.-I. Skullerud, \emph{Ab initio calculation of finite-temperature charmonium potentials}, Phys. Rev. D \textbf{89} (2014) 071502. 

\bibitem{chen} Jiao-Kai Chen, \emph{Spectral method for the Cornell and screened Cornell potentials in momentum space}, Phys. Rev. D \textbf{88} (2013) 076006. 

\bibitem{hamz} M. Hamzavi and A. A. Rajabi, \emph{Scalar-vector-pseudoscalar Cornell potential
for a spin-1/2 particle under spin and
pseudospin symmetries: 1+1 dimensions}, Ann Phys-New York \textbf{334} (2013) 316 - 320.

\bibitem{claudio} Claudio O. Dib and Nicol\'as A. Neill, \emph{$\chi_{b}(3P)$ splitting predictions in potential models,} Phys. Rev. D \textbf{86} (2012) 094011.

\bibitem{bali} Gunnar S. Bali, \emph{QCD forces and heavy quark bound states}, Phys. Rep. \textbf{343} (2001) 1.

\bibitem{kang} D. Kang and E. Won, \emph{Precise numerical solutions of potential problems using the Crank–Nicolson method}, J. Comput. Phys. \textbf{20} (2008) 2970.

\bibitem{hall7} R. L. Hall, \emph{ Simple eigenvalue formula for the Coulomb-plus-linear potential,}  Phys. Rev. D \textbf{30} (1984) 433.

\bibitem{aim} H. Ciftci, R. L Hall and N. Saad, \emph{Asymptotic iteration method for eigenvalue problems},  J. Phys. A: Math. Gen. 36 (2003) 11807.

\bibitem{hall1}R. L. Hall, {\it Energy trajectories for the N-boson problem by the method of potential envelopes}, Phys. Rev. D {\bf 22}, 2062 (1980).

\bibitem{hall3}R. L. Hall, {\it A geometrical theory of energy trajectories in quantum mechanics}, J. Math. Phys. {\bf 24}, 324 (1983).

\bibitem{hall4}R. L. Hall, {\it Kinetic potentials in quantum mechanics}, J. Math. Phys. {\bf 25}, 2078 (1984).

\bibitem{hall5}R. L. Hall, {\it Spectral geometry of power-law potentials in quantum mechanics}, Phys. Rev. A {\bf 39}, 5500 (1989).

\bibitem{hall6}R. L. Hall, {\it Envelope theory in spectral geometry}, J. Math. Phys. {\bf 34}, 2779 (1993).


\bibitem{atkin} K. Atkinson and W. Han, {\it Spherical harmonics and approximations on the unit sphere: An introduction}  (Springer, New York, 2012).

\bibitem{doren1986} D. J. Doren and D. R. Herschbach, \emph{Inter-dimensional degeneracies, near degeneracies and their applications}, J. Chem. Phys. \textbf{85} (1986) 4557.

\bibitem{simon} M. Reed and B. Simon, \emph{
Methods of Modern Mathematical Physics, IV. Analysis of Operators} , Academic Press, New York, 1978. The appropriate 
discrete-spectrum result for the linear-plus-Coulomb potential is given by Theorem XIII.69 on p~250.

\bibitem{abr} M. Abramowitz and I. Stegun, \emph{Handbook of mathematical functions},
Dover Publications: New York (1965).

\bibitem{landau} L. D. Landau and E. M. Lifshitz, \emph{Quantum Mechanics: non-relativistic theory,} Pergamon, London, 1981.

  





 





\end{thebibliography}
\end{document}